\documentclass[12pt]{article}
\pdfoutput=1
\usepackage[latin1]{inputenc}

\usepackage{amsmath}
\usepackage{amsfonts}
\usepackage{amssymb}
\usepackage{graphicx}
\usepackage{geometry}
\usepackage{amssymb,epsfig,subfigure}
\usepackage{hyperref}

\def\simge{
    \mathrel{\rlap{\raise 0.511ex 
        \hbox{$>$}}{\lower 0.511ex \hbox{$\sim$}}}}

\def\simle{
    \mathrel{\rlap{\raise 0.511ex 
        \hbox{$<$}}{\lower 0.511ex \hbox{$\sim$}}}}

\makeatletter
\renewcommand\section{\@startsection {section}{1}{\z@}%
                                 {-3.5ex \@plus -1ex \@minus -.2ex}
                                   {2.3ex \@plus.2ex}%
                                   {\normalfont\large\bfseries}}
\renewcommand\subsection{\@startsection{subsection}{2}{\z@}%
                                   {-3.25ex\@plus -1ex \@minus -.2ex}%
                                     {1.5ex \@plus .2ex}%
                                     {\normalfont\bfseries}}
\renewcommand\subsubsection{\@startsection{subsubsection}{3}{\z@}%
                                   {-3.25ex\@plus -1ex \@minus -.2ex}%
                                     {1.5ex \@plus .2ex}%
                                     {\normalfont\itshape}}
\makeatother

\def\pplogo{\vbox{\kern-\headheight\kern -29pt
\halign{##&##\hfil\cr&{\ppnumber}\cr\rule{0pt}{2.5ex}&\ppdate\cr}}}
\makeatletter
\def\ps@firstpage{\ps@empty \def\@oddhead{\hss\pplogo}%
  \let\@evenhead\@oddhead 
}
\def\maketitle{\par
 \begingroup
 \def\thefootnote{\fnsymbol{footnote}}
 \def\@makefnmark{\hbox{$^{\@thefnmark}$\hss}}
 \if@twocolumn
 \twocolumn[\@maketitle]
 \else \newpage
 \global\@topnum\z@ \@maketitle \fi\thispagestyle{firstpage}\@thanks
 \endgroup
 \setcounter{footnote}{0}
 \let\maketitle\relax
 \let\@maketitle\relax
 \gdef\@thanks{}\gdef\@author{}\gdef\@title{}\let\thanks\relax}
\makeatother

\numberwithin{equation}{section}

\renewcommand{\dag}{\dagger}

\newcommand{\be}{\begin{equation}}
\newcommand{\bea}{\begin{eqnarray}}
\newcommand{\ee}{\end{equation}}
\newcommand{\eea}{\end{eqnarray}}
\newcommand\beq{\begin{equation}}
\newcommand\eeq{\end{equation}}

\newcommand{\tr}{{\rm tr}}

\def\be{\begin{equation}}
\def\ee{\end{equation}}
\def\ba#1\ea{\begin{align}#1\end{align}}
\def\bg#1\eg{\begin{gather}#1\end{gather}}
\def\bm#1\em{\begin{multline}#1\end{multline}}
\def\bmd#1\emd{\begin{multlined}#1\end{multlined}}

\def\k{\kappa}
\def\l{\lambda}

\def\y{\psi}

\def\fr{\frac}
\def\na{\nabla}
\def\pa{\partial}

\def\({\left(}
\def\){\right)}
\def\[{\left[}
\def\]{\right]}

\begin{document}

\setcounter{page}0
\def\ppnumber{\vbox{\baselineskip14pt
}}
\def\ppdate{\footnotesize{SLAC-PUB-15909, SU-ITP-14/06}} \date{}

\author{Xi Dong, Samuel McCandlish, Eva Silverstein, Gonzalo Torroba\\
[7mm]
{\normalsize \it Stanford Institute for Theoretical Physics }\\
{\normalsize  \it Department of Physics, Stanford University }\\
{\normalsize \it Stanford, CA 94305, USA}\\
[3mm]
{\normalsize \it Theory Group, SLAC National Accelerator Laboratory}\\
{\normalsize  \it Menlo Park, CA 94025, USA}\\
[3mm]}

\bigskip
\title{\bf  Controlled non-Fermi liquids from \\ spacetime dependent couplings
\vskip 0.5cm}
\maketitle

\begin{abstract}
We construct perturbatively controlled non-Fermi liquids in 3+1 spacetime dimensions, using mild power-law translation breaking interactions. Our mechanism balances the leading tree level effects from such gradients against quantum effects from the interaction between the Fermi surface and a critical boson. We exhibit this in a model where finite density fermions interact with a scalar field via a Yukawa coupling of the form $g(x)\propto |x|^\kappa$. The approximate non-Fermi liquid behavior arises in the limit of small $\kappa$ and persists over an exponentially large window of scales, being cut off by the regime where the coupling becomes large, or by superconducting instabilities.  The translation breaking coupling introduces anisotropic deformations of the Fermi surface depending on the direction of the gradient.    
An extension of this mechanism to 2+1 dimensions could provide a strongly translation-breaking, but weakly coupled non-fermi liquid, something we leave for further work.
\end{abstract}
\bigskip
\newpage

\tableofcontents

\vskip 1cm

\section{Introduction}\label{sec:intro}

Many modern condensed matter problems involve a combination of strong dynamics and finite density, making their study very interesting and also extremely challenging. An important goal is then to develop tools that can provide analytic insight on their underlying physics.  One basic class of problems involves the so-called non-Fermi liquids, systems which are thought to have a Fermi surface but which exhibit thermodynamic and transport properties distinct from those which can be produced by weakly interacting fermions.      
One fruitful approach along these lines has been to identify special limits of parameters for which the physics is under analytic control, and then try to extend the analysis beyond the original range of applicability. Famous examples include the epsilon expansion and the large $N$ limit. 

In this work we present another mechanism that leads to controlled critical phases of potential relevance to condensed matter systems. The basic idea -- developed in \cite{Dong:2012ua}\ for relativistic theories -- is to deform a theory with a marginal coupling $g$ by adding mild space or time dependence, $g(x) =g_0 |x|^\kappa$, with $\kappa$ a small number. Such dependence can be obtained physically by realizing the coupling $g(x)$ as the background value of a field.  Balancing the classically relevant scaling contribution $|x|^\kappa$ against quantum corrections can give rise to an approximate fixed point perturbatively in $\kappa$. This leads ~\cite{Dong:2012ua}\ to a variety of new critical phenomena: an analogue of the small-$\epsilon$ Wilson-Fisher fixed point in a physical (integer) dimensionality, new fixed points in four-dimensional gauge theories and various other examples in two- and three-dimensional field theories. 

Our goal here is to extend this to systems at finite density, specifically the problem of controlling non-Fermi liquids.   This problem is motivated by experimental results on materials such as high-$T_c$ superconductors, and it has been attacked in a variety of ways theoretically over the years.  One important recent development was the realization \cite{lee2}\ that the large-$N_\text{flavor}$ approximation does not control the problem because of certain infrared effects on the Fermi surface.   Although one can obtain insight from further analysis of this strongly coupled system \cite{SachdevMet}, it is useful to approach the problem by introducing a new control parameter.  In the works  \cite{NW1,NW2,Mross:2010rd}, for example, a bosonic action non-analytic in momentum can be chosen in such a way as to produce a dynamical critical exponent $z$ close to the value $z_c=2$ for which the coupling is marginal.  Another parameter one can formally introduce is $\epsilon$, a small deviation from integer dimensionality, in order to produce a nearly marginal classical coupling  and obtain a weakly coupled fixed point analogous to the small-$\epsilon$ Wilson Fisher theory.  With the aid of an additional expansion in large $N$ one can extend this to a physical dimensionality with $\epsilon= 1$ 
\cite{NFLepsilon}.  Such an $\epsilon$ expansion can be done in various ways, depending on which directions are affected by the shift in dimensionality \cite{leeD}.  

For small $\epsilon$, this is a formal procedure (since the dimensionality is not an integer in that case).   For small $z-z_c$, the physical origin of the new parameter is also not completely clear, although it is fair to say that non-analytic effective actions can sometimes arise effectively, one example being at the infrared fixed point in three-dimensional (relativistic) QED \cite{appelquist}.  

In this paper, we will introduce an arguably more physical control parameter, obtaining a nearly marginal coupling via a small translation breaking effect coming from spatially varying couplings which classically scale as a power $\sim |x|^\kappa$, $\kappa\ll 1$.  Our main result will be a non-Fermi liquid in $3+1$ dimensions, perturbatively controlled in the presence of the small gradient parameter $\kappa$.  Since the dimensionality of spacetime is integer, and the spatially varying coupling can in principle be obtained as a background profile of another field, this setup seems somewhat easier to obtain physically.  Possible sources of translation breaking include defects and impurities found in materials.  In any case, this provides a new method for approaching the problem, one which may tie into other aspects of translation-breaking effects in finite density field theory and condensed matter.

\section{Non-Fermi liquid from spacetime dependent couplings}\label{sec:NFLphi}

We will analyze the theory of a spherical Fermi surface interacting with critical bosons in $3+1$ dimensions ($3$ spatial dimensions), with the following Euclidean action
\bea\label{eq:L1}
S &=&\int d^4 x \;\Bigg\{  \fr12 \,\tr\left((\partial_\tau \phi)^2+c^2 (\vec \nabla \phi)^2\right) +\frac{\lambda(x)}{N^2} \left(\tr(\phi^2)\right)^2+\frac{\rho(x)}{N} \tr(\phi^4)\nonumber\\
&& \qquad \qquad+ \psi_i^\dag \(\pa_\tau -\fr{\vec\na^2}{2m} - \mu_F\) \psi_i + \frac{g(x)}{\sqrt{N}} \phi_{ij} \y_i^\dag \y_j \Bigg\}\,.
\eea
The model has an $SU(N)$ global symmetry under which the fermions are in the fundamental representation and the boson is in the adjoint. In this first part of our analysis it will be convenient to take the large $N$ limit, which simplifies some of the quantum corrections, by suppressing the backreaction of the fermions on the scalar sector or the generation of $\psi^4$ interactions that could lead to instabilities. Later, in \S \ref{sec:smallN}, we will incorporate finite $N$ effects and argue that our conclusions are also valid away from the large $N$ expansion. Furthermore, in order to obtain a nontrivial critical point for the boson, it will be enough to keep only one of the scalar quartic couplings, so in what follows we set $\rho(x)=0$. 

The couplings $\lambda(x)$ and $g(x)$, which are kept finite at large $N$, have a mild space-time dependence of the form
\be\label{eq:gradients1}
\lambda(x) = \lambda_0\,| x|^{\kappa_1}\;,\;g(x) = g_0 |x|^{\kappa_2}\,,
\ee
where $0 <\kappa_i \ll 1$. The subindex `$0$' refers to bare couplings; we will shortly discuss how to define a renormalized theory at a scale $\mu$, introducing physical dimensionless couplings that we will denote by $\lambda$ and $g$.
The couplings could depend on space, on time or on some isotropic or anisotropic combination; for now we use $|x|$ to mean any of these possibilities, and below we will study in more detail the effects of isotropic or anisotropic gradients. In specific realizations of these gradients the exponents $\kappa_1$ and $\kappa_2$ may be related (for instance, in some models it is natural to have $\lambda (x) \sim g(x)^2$), but for now we will keep them as arbitrary small numbers.

From~\cite{Dong:2012ua}, the purely bosonic theory with $\l(x)\phi^4$ interaction, consisting of the first two terms in the above action, has an approximate scale-invariant fixed point with $\l \sim\k_1/N^2$.  The same is true for the full action (\ref{eq:L1}) at large $N$, since in this limit quantum corrections from fermions are subleading. We will review this shortly in \S \ref{subsec:scalar}. On the other hand, we will find in \S \ref{subsec:fermion} that the Yukawa interaction produces important modifications on the fermionic sector, leading to a non Fermi liquid under analytic control at small $\kappa_2$. Before turning to this analysis, let us discuss the classical scalings of the theory.

\subsection{Classical scaling analysis}\label{subsec:scaling}

The renormalization group analysis of interacting theories at finite density \cite{FEFT}\ is complicated by the fact that bosons and fermions have very different low energies degrees of freedom---while the former have light excitations only near the origin of momentum space, the latter have a whole Fermi surface of low energy degrees of freedom. We regularize the theory using
dimensional regularization, which is an efficient scheme particularly in multiloop calculations or for theories with gauge fields. In this approach one 
 continues the integrals from four to $D=4-\epsilon$ dimensions ($d=3-\epsilon$ spatial dimensions), and then takes $\epsilon \to 0$. In a minimal subtraction scheme, we subtract pole contributions, which has the effect of replacing $\frac{1}{\epsilon} \to \log \frac{\Lambda_{UV}}{\Lambda_{IR}}$, where $\Lambda_{UV,IR}$ are related to the renormalization scale and the energy scale of a process.
In a Wilsonian scheme, these scales correspond to the UV cutoff scale, and the scale down to
which one integrates out shells of modes. 
We stress that the theory is always in a physical number $D=4$ of dimensions:  $\epsilon$ is used as a regulator that is taken to zero at the end of the calculation. As in~\cite{Dong:2012ua}, the nontrivial critical phenomena will come from the small $\kappa$ effects.

An important aspect of the RG is that momenta of
bosons and fermions are scaled differently.  Here we will follow the scaling procedure recently discussed in~\cite{NFLepsilon}, where the former are scaled towards the origin, while the latter are scaled towards the Fermi surface. To see this in more detail, split the momenta into components that are perpendicular and parallel to a given unit vector $\hat n$ normal to the spherical Fermi surface:
\be
\vec k = \hat n \left(\ell + k_\text{F}\right)
\ee
The low energy limit for fermions is $\ell \ll k_F=\sqrt{2m \mu_F}$; $\hat n$ describes angles tangential to the surface.
For fermions only $\omega$ and the distance to the Fermi surface are scaled, 
\be
\omega'= e^b \omega\;,\; \ell' = e^b  \ell\;
\ee
In particular, in this `spherical' RG the fermion dispersion relation is $\omega = v \, \ell$, where $v=k_F/m$. Bosons have the usual relativistic scaling of all momentum components
\be\label{eq:boson-scaling}
k_\mu ' = e^b k_\mu\,.
\ee

Given these scalings, the dimensions of the momentum-space fields near the free field theory limit are, in $D=4-\epsilon$ dimensions,
\be
\phi'(k_\mu') = e^{-(3- \epsilon/2)b} \phi(k)\;,\;\psi(k_\mu') =e^{- \frac{3}{2}b} \psi(k)
\ee
Finally, we come to the scaling of the spacetime dependent couplings. Since we may view $\lambda(x)$ and $g(x)$ as background values of additional bosonic fields, it is natural to scale the momentum dependence of the Fourier-transformed couplings as in (\ref{eq:boson-scaling}). Doing so gives
\be
\lambda_0' = e^{(\epsilon+\kappa_1)b} \lambda_0\;,\;g_0' =e^{ (\epsilon/2+ \kappa_2)b}g_0\,.
\ee

\begin{figure}
\begin{centering}
\includegraphics[scale=0.3]{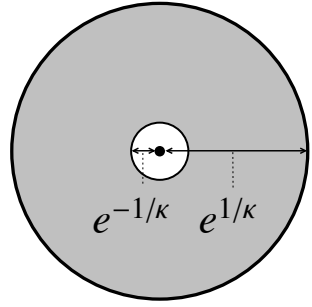}\protect
\caption{\small{For $g(x)$ or $\lambda(x) \sim |\mu^2 \left(x^2 + y^2 + z^2\right)|^{\kappa/2}$, our analysis is valid in the exponentially large shaded spatial region shown above.  The center point is the point towards which the spatial coordinates are scaling, which physically could represent a defect, impurity, or other point source in a material.  An analogous picture holds for $g(x) \sim |\mu^2 \left(x^2 + y^2\right)|^{\kappa/2}$ or $|\mu z|^{\kappa}$, with the central point replaced by a line or a plane respectively.}}
\end{centering}
\end{figure}

\begin{figure}
\begin{centering}
\includegraphics[scale=0.8]{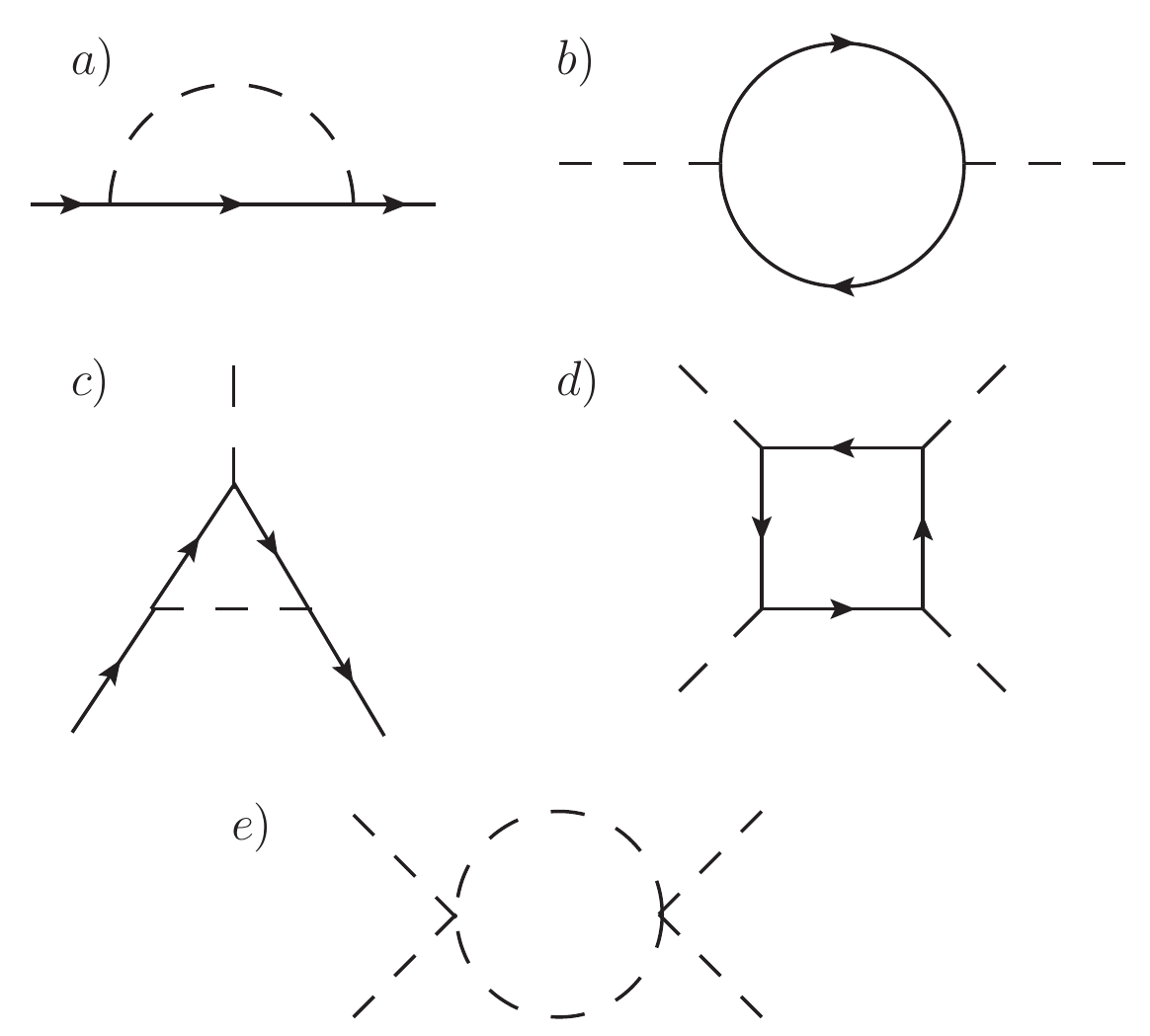}\protect
\caption{\small{These one-loop diagrams are required for our analysis.  a) Fermion self-energy.  b) Boson self-energy. c) Yukawa interaction. d) Correction to the boson quartic interaction due to a fermion loop.  e) Boson quartic correction.}}
\end{centering}
\end{figure}

\subsection{Bosonic sector}\label{subsec:scalar}

Having specified the RG procedure and the scaling of fields, let us analyze the quantum effects from one-loop corrections first in the bosonic sector and then for fermions. As we discussed before, at large $N$ the effects of fermions on the renormalization of the bosonic couplings is negligible. Therefore the scalar part of our theory is as in~\cite{Dong:2012ua}, leading to an approximate perturbative fixed point valid over a parametrically large window of scales. We now briefly review this mechanism, and refer the reader to that reference for more details.

The basic idea is that for for $0< \kappa_1 \ll 1$, the gradient gives the leading contribution to the classical running of the coupling, while its effects on the one loop corrections can be neglected over a wide range of scales
\be\label{eq:windowkappa1}
e^{-1/\kappa_1}\ll |\mu x| \ll e^{1/\kappa_1}\,.
\ee
This can be seen most clearly from the one loop correction to the quartic coupling in position space (ignoring for now the index structure of the fields):\footnote{In this section we set $c=1$ for simplicity of presentation; it is easy to reintroduce $c$ based on dimensional analysis.}
\be\label{eq:lambda-oneloop}
S_\text{eff}\supset\,\int d^Dx\,d^Dx' \,\phi(x)^2 \,\frac{\lambda(x)\lambda(x')}{|x-x'|^{2(D-2)}} \phi(x')^2\,.
\ee
For small $\kappa_1$ the one loop correction is dominated by the UV region $x_-=x-x' \to 0$ and in this limit the couplings are approximately independent of $x_-$. Therefore the one loop correction is the same as in the theory with constant interactions. We also recall that at one loop there is no wavefunction renormalization, so the boson propagator reads
\be
G_B(x,x') =  \int \frac{d\omega \,d^{d}k}{(2\pi)^D}\,\frac{e^{-i \omega \tau+i \vec k \cdot \vec x}}{\omega^2+ \vec k^2}= \frac{1}{4\pi^2|x-x'|^2}\,.
\ee
Furthermore, the mass is fine-tuned to vanish.

Let us introduce the renormalized dimensionless coupling $\lambda$, which is defined as the value of the scalar four-point function at the scale $\mu$ (after amputating the external legs):
\be\label{eq:lambdaphys}
G^{(4)}_{amp}=- \mu^\epsilon  |\mu x|^{\kappa_1}\,\lambda\;,\;{\rm at}\;\;s=t=u=\mu^2\,.
\ee
Evaluating the one loop contribution (\ref{eq:lambda-oneloop}) using dimensional regularization shows that the bare coupling has to be chosen to be (at leading order in $1/N$)
\be
 \lambda_0 \approx \mu^{\epsilon+\kappa_1}\, \lambda\left( 1+ \frac{3\lambda^2}{16 \pi^2} N^2 \frac{1}{\epsilon} + \ldots\right)
\ee
so that the pole in $1/\epsilon$ cancels and (\ref{eq:lambdaphys}) is obtained. Since the bare coupling is independent of the RG scale, this gives a beta function
\be
\beta_{\lambda} = \mu \frac{\partial \lambda}{\partial\mu}= -(\epsilon+\kappa_1)\lambda + \frac{3\lambda^2}{16 \pi^2} N^2 + \ldots
\ee
which implies an approximate fixed point
\be
\lambda \approx \frac{16 \pi^2}{3N^2} \kappa_1
\ee
in the limit $\epsilon \to 0$.

\subsection{Fermionic sector}\label{subsec:fermion}

We now consider the quantum corrections to the Fermi surface from interactions with the approximate fixed point for the scalar.
At one-loop order, the only new diagrams correspond to the correction to the fermion self-energy, because the Yukawa vertex correction and boson self-energy are suppressed by $1/N$.

Let us first show that for small $\kappa_2$ the one loop answer is well-approximated by the translationally-invariant limit. Working in position space and summing the geometric series gives the one loop inverse propagator
\be
(G_F^{(1)})^{-1}(x,x')= (G_F^{(0)})^{-1}(x,x') -\Sigma(x,x')\,.
\ee
Here $G_F^{(0)}$ is the tree level fermion propagator,
\be
G_F^{(0)}(x,x')=- \int \frac{d\omega\,d^{d}k}{(2\pi)^D}\,\frac{e^{-i \omega \tau+i \vec k \cdot \vec x}}{i \omega- \frac{\vec k^2}{2m}+\mu_F}
\ee
and the one-loop fermion self-energy is given by
\be\label{eq:Sigma1}
\Sigma(x, x')=g(x) G_B^{(0)}(x,x')G_F^{(0)}(x,x')g(x')\,.
\ee

For small $\kappa_2$ the contribution of $\Sigma$ to the effective action is dominated by $x_- \to 0$. This means that the couplings are approximately independent of $x_-$ and the answer is the same as in the translationally-invariant theory.
As in \cite{Dong:2012ua}, in order to understand the small-$\kappa$ analogue of the small $\epsilon$ fixed point of \cite{NFLepsilon}, at leading order in $\kappa$ we will only need the translationally-invariant limit of the one-loop fermion self-energy.  

We now describe in some detail the renormalization procedure in this theory, which is a bit more involved than in relativistic QFT. Focusing on the fermionic part of the theory, the bare Lagrangian is written in terms of renormalized couplings and fields, plus counterterms:\footnote{See e.g.~\cite{Peskin:1995ev}.}
\be\label{eq:Lrenorm}
\mathcal{L}=\psi_{R}^{\dag}\bigl(\left(1+\delta_{Z}\right)\partial_{\tau}+\left(v+\delta_{v}\right)\partial_{\perp}\bigr)\psi_{R}+\frac{g}{\sqrt{N}}\left(\mu^{\epsilon/2+\kappa_{2}}+\delta_{g}\right)|x|^{\kappa_{2}}\phi_{R}\psi_{R}^{\dag}\psi_{R}\,.
\ee
The fermion propagator has been linearized around the Fermi surface, and $\partial_\perp$ is the derivative perpendicular to it.
The bare and renormalized fields are related by $\psi = Z_\psi^{1/2} \psi_R$ and $\phi= Z_\phi^{1/2} \phi_R$ (recall that at one loop $Z_\phi=1$, so this factor will not enter the following analysis). The additional terms in (\ref{eq:Lrenorm}) are the counterterms, defined as
\be
\delta_Z = Z_\psi-1\;,\;\delta_v = Z_\psi v_0- v\;,\;g \delta_g = g_0 Z_\psi Z_\phi^{1/2} - g \mu^{\kappa_2+\epsilon/2}\,.
\ee

Due to the breaking of Lorentz invariance, the time and spatial components of the kinetic term can renormalize independently, so we have allowed for the wavefunction counterterm $\delta_Z$ as well as a renormalization of the Fermi velocity $\delta_v$.

Next we have to calculate the fermion two-point function at one loop using the renormalized Lagrangian (\ref{eq:Lrenorm}). Summing the geometric series of corrections gives an inverse propagator
\be
-G_F^{-1}(\omega, \ell) =( i \omega - v \ell)+(i \omega \delta_Z - \delta_v \ell)+ \Sigma(\omega,\ell)\,,
\ee
where the one-loop contribution $\Sigma$ to the self-energy $(i \omega \delta_Z - \delta_v \ell)+ \Sigma(\omega,\ell)$ is given by
\be\label{eq:Sigma2}
\Sigma(\omega_e,\,\ell_e) =- g^2\mu^\epsilon \int \frac{d \omega dl d^{d-1} k_\parallel}{(2\pi)^D}\,\frac{1}{\omega^2 +c^2( \ell^2 + \vec k_\parallel^2)}\,\frac{1}{i(\omega+\omega_e) - v(\ell+\ell_e)}\,.
\ee
Here (\ref{eq:Sigma2}) is the Fourier transform of (\ref{eq:Sigma1}) in the limit of small $\kappa_2$ and 
\be\label{eq:exp2}
e^{-1/\kappa_2}\ll |x \mu| \ll e^{1/\kappa_2}\,,
\ee
so that we can neglect the one-loop effects from gradients.

The calculation of the self-energy is somewhat involved and is shown explicitly in the Appendix; see also~\cite{NFLepsilon}. The result is
\be\label{poleplusfinite}
\Sigma(\omega, \ell) = \frac{\Sigma_{-1}}{\epsilon} + \Sigma_0 + \mathcal O(\epsilon)
\ee
with
\be
\Sigma_{-1}= \frac{g^{2}}{4\pi^{2}c^{2}\left(c+\left|v\right|\right)}\left(i\omega+\text{sign}\left(v\right) c \ell\right)
\ee
and
\begin{eqnarray}\label{Sigmazero}
\Sigma_{0}
 & = & \frac{g^{2}}{8\pi^{2}c^{2}\left(c^{2}-v^{2}\right)}\bigg(2c\left(v\ell-i\omega\right)\log\left(\frac{v\ell-i\omega}{\mu}\right)-(c+v)\left(c\ell-i\omega\right)\log\left(\frac{c\ell-i\omega}{\mu}\right)\nonumber\\
 &&-(c-v)\left(c\ell+i\omega\right)\log\left(\frac{c\ell+i\omega}{\mu}\right)\bigg)+ \ldots
 \end{eqnarray}
 The extra terms not shown in $\Sigma_0$ are linear in $\omega$ and $\ell$, and do not contain logs, so they do not contribute to the RG.  

We are now ready to subtract the divergences and compute the beta functions. 
A convenient formal prescription is minimal subtraction, in which we simply subtract the pole contribution in 
(\ref{poleplusfinite}).  
This fixes the counterterms
\be
\delta_Z = - \frac{g^2}{4\pi^2 c^2 (c+\left|v\right|)} \frac{1}{\epsilon},\; ~~~ \delta_v =\frac{g^2 \text{sign}\left(v\right)}{4\pi^2 c (c+\left|v\right|)} \frac{1}{\epsilon} 
\ee
The renormalization of the Yukawa coupling is of order $1/N$, so we will neglect it at large $N$.  In this section, we will analyze our theory using the simplifications of large $N$, and thus set $\delta_g=0$.  Later we will relax this requirement and analyze the vertex correction and other finite $N$ features of our system.  

Finally, requiring that the bare couplings be independent of the RG scale gives the beta functions for the running couplings\footnote{Recall that the beta functions can also be read off from the Callan-Symanzik equations. These have an extra term $\beta_v \frac{\partial}{\partial v}$ as compared to their relativistic version, due to the independent running of the Fermi velocity.}
\bea\label{eq:RG1}
\gamma&=& \frac{1}{2}\,\frac{d \delta_Z}{d \log \mu} = \frac{1}{2}\,\frac{\partial \delta_Z}{\partial g}\frac{\partial g}{\partial \log \mu} =\frac{g^2}{8 \pi^2 c^2 (c+\left|v\right|)}\nonumber\\
\beta_v &=& - \frac{d \delta_v}{d \log \mu}+2v \gamma=\frac{g^2}{4\pi^2 c^2}\text{sign}\left(v\right) \\
\beta_g&=&g \left(-\left(\frac{\epsilon}{2}+ \kappa_2 \right) - \frac{\partial \delta_g}{\partial \log \mu}+2 \gamma\right)=g\left(-\frac{\epsilon}{2}- \kappa_2 + \frac{g^2}{4 \pi^2 c^2 (c+\left|v\right|)}\right)\nonumber\,.
\eea
These are the main results of our analysis, and now we will discuss their implications for the physical theory in $3+1$ dimensions ($\epsilon=0)$.

The second and third equations in (\ref{eq:RG1}) give a coupled system for the RG evolution of $v$ and $g$. The beta function $\beta_v$ implies that $v$ decreases towards the IR. The beta function for $g$ has two contributions: a tree-level term from the spacetime dependent coupling, which makes it relevant, and a quantum correction which tries to make $g$ irrelevant. Solving this system of equations shows that there is a scale $\mu_*$ at which the Fermi velocity vanishes and then stops running. Below this scale, $\beta_g$ admits a perturbative fixed point at
\be
g^2 \approx 4\pi^2 c^3 \kappa_2\,,
\ee  
in the limit $\epsilon \to 0$, namely in 3+1 dimensions. Higher loop corrections are negligible for small $\kappa_2$.
This fixed point is valid over an exponential window of scales (\ref{eq:exp2}).

At this approximate fixed point, the fermion acquires a nontrivial anomalous dimension controlled by $\kappa_2$,
\be
\gamma \approx \frac{g^2}{4\pi^2c^3}\,\approx\,\kappa_2\,.
\ee
Since $v \to 0$ as we approach the critical point, the dependence on the momentum $\ell$ disappears and
the fermion two-point function becomes
\be\label{eq:GNFL}
G_F(\omega) \approx \frac{i}{ \omega^{1- \gamma}}\,.
\ee
It exhibits non-Fermi liquid behavior, with $\epsilon=0$ and small $\k_2$, in a controlled perturbative approximation scheme. It is possible that a higher order dependence in $\ell$ is generated by subleading quantum corrections, which would signal a transition to a theory with higher dynamical exponent $z$. Another intriguing possibility would be if all momentum dependence becomes irrelevant as we approach the fixed point, which would lead to a local quantum critical phase with correlation function (\ref{eq:GNFL}). It will be very interesting to understand the fate of the momentum dependence in detail.\footnote{Consequences of $v \to 0$ in models in $d=3-\epsilon$ dimensions are studied in~\cite{shamitfuture}.}


\begin{figure}
\begin{centering}
\includegraphics[scale=0.3]{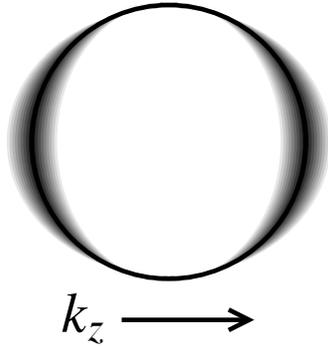}\protect\caption{\small{A gradient in the $\hat{z}$ direction does not affect points on the Fermi surface perpendicular to $\hat{z}$, but other zero-energy states no longer have definite $k_z$.}}
\end{centering}
\end{figure}

\subsection{Effect of gradients on the Fermi surface}\label{subsec:anisotropies}

So far we worked at the leading order in our small parameter $\kappa_2$.  This sufficed to exhibit non-Fermi liquid behavior over a large range of scales; in particular, we computed a frequency-dependent contribution to the fermion self energy going like $\omega \log\omega$.

It is also interesting to consider the effect of spatial translation breaking on the geometry of the Fermi surface. 
The translation-breaking perturbation of the Hamiltonian changes the spectrum of Fermion levels in general.  We will briefly explore that in this section.  To be specific, let us consider a gradient in one spatial direction, labeled $z$, i.e.  
\be\label{gz}
g(x)=g(z)=g_0|\mu z|^{\kappa_2}\,.
\ee
We will denote the other spatial coordinates $\vec x_\perp$.  

The fermionic excitations through one-loop order satisfy
\be\label{Feom}
\int d^D x'\left(\delta(x-x')[\partial_\tau -\frac{\vec\nabla^2}{2m}-\mu_F] - g(z)\Sigma (x-x') g(z')\right) \psi(x') =0\,,
\ee
with $\Sigma$ the one-loop fermion self-energy in the homogeneous theory.  
Let us first Fourier transform this in the translationally-invariant $\tau, \vec x_\perp$ directions, giving
\be\label{Feompartk}
\int dz'\left(\delta(z-z')[-\omega + \frac{\vec k_\perp^2}{2 m} -\frac{\partial_z^2}{2m}-\mu_F]+ g(z)\Sigma_{\omega, \vec k_\perp} (z-z') g(z') \right) \psi_{\omega, \vec k_\perp} (z')=0\,.
\ee 
Without the one-loop correction $\Sigma$, we recover the usual Fermi surface, satisfying (\ref{Feompartk}) with a momentum eigenstate $\psi_{\omega, \vec k_\perp} (z)=e^{ik_z z}$ and $\vec k^2=\vec k_\perp^2+k_z^2=2m\mu_F$.    

With the translation-breaking deformation, it is simple to check that momentum eigenstates do not survive as solutions:  again putting in the ansatz  $\psi_{\omega, \vec k_\perp} (z)=e^{ik_z z}$ we obtain (at $\omega=0$)
\be\label{fulleqnmom}
 \frac{\vec k_\perp^2}{2 m} +\frac{k_z^2}{2m}-\mu_F+g(z)\int dz' \Sigma_{\omega, \vec k_\perp}(z-z')g(z')e^{i k_z z'} = 0.
\ee
This can only be satisfied if the integral in the second term is either zero or proportional to $1/g(z)$.  We can exclude this possibility as follows.      

From (\ref{Sigmazero}) we find that $\Sigma$ (the one-loop Fermion self-energy at the level of the homogeneous theory) depends only on the distance $\ell$ from the Fermi surface in momentum space, going like $\ell \log \ell$.  Let us again write $\vec k=(\vec k_\perp, p_z) = \hat n\left(k_\text{F} + \ell \right)$ (with $ k_F \hat n$ a point on the original Fermi surface and $\ell\ll k_F$).  Then 
\be\label{Sigzl}
\Sigma_{\omega=0, \vec k_\perp}(z-z')\sim \int dp_z e^{ip_z(z-z')}(-c^2 \ell \log \frac{c|\ell|}{\mu}).
\ee
The dependence of $\ell$ on $p_z$ depends on the direction of $\vec k_F$ here.  If $\hat{n} \propto \vec k_\perp$, i.e. orthogonal to the $z$ direction, then $\ell$ is independent of $p_z$  In this direction, we recover the usual Fermi surface for $\ell=0$.  
In the case where $\hat n$ points along the $z$ direction, we have $p_z=k_F+\ell$.  The integral (\ref{Sigzl}) becomes
\be\label{Sigzlagain}
\int d\ell \,v e^{i(k_F+\ell)(z-z')}(-c^2 \ell \log \frac{c|\ell|}{\mu})
\ee
The Fourier Transform of $\ell \log|\ell|$ is 
\be\label{FT}
-i\sqrt{\frac{\pi}{2}}\frac{{\rm sgn}(z-z')}{(z-z')^2}.
\ee
Plugging this into the last integral in (\ref{fulleqnmom}), we find a rich function of $z$,
not zero or a result proportional to $1/g(z)$.  

Thus, the original Fermi surface is modified by our inhomogeneous coupling.  It cannot be described as a locus in momentum space, since the corrected Fermion energy eigenstates are not momentum eigenstates in the $z$ direction.  In the $\vec x_\perp$ directions, they remain momentum eigenstates, so this modification of the Fermi surface (or more generally the space of Fermion ground states) is anisotropic.  It would be interesting to understand if this could have any connection to Fermi surface anisotropies (such as hot and cold spots) in condensed matter physics.

\section{Dynamics at finite $N$}\label{sec:smallN}

In the previous section we argued that, at large $N$, the theory (\ref{eq:L1}) leads to a non Fermi liquid interacting with a critical $\phi^4$ scalar theory in $3+1$ dimensions. These approximate critical phenomena arise from gradients (\ref{eq:gradients1}) in the scalar and Yukawa couplings, and are valid over an exponentially large window of scales $e^{-1/\kappa}\ll |\mu x| \ll e^{1/\kappa}$. The large $N$ limit allowed us to obtain the fixed point in a simple way, by suppressing effects like the backreaction of fermions into the scalar sector. Now we will study the theory at finite $N$, which is important for trying to realize our mechanism in realistic systems.

\subsection{Incorporating fermion backreaction}\label{subsec:backr}

Let us first incorporate the effects of fermion quantum corrections on the scalar sector. They renormalize the scalar wavefunction but not the $\phi^4$ vertex, which we now calculate.

In the limit of small $\kappa_2$ the one loop boson self-energy from the finite density fermions reduces to the translationally invariant result,
\be\label{eq:Pi1}
\Pi(p_0, \vec p)= \frac{g^2}{N} \int \frac{d\omega\,d^d k}{(2\pi)^{d+1}} \,\frac{1}{i \omega - \frac{\vec k^2}{2m}+\mu_F}\, \frac{1}{i (\omega+p_0) - \frac{(\vec k+\vec p)^2}{2m}+\mu_F}\,.
\ee
By power-counting we would naively expect $\Pi \sim 1/\epsilon$ at small $\epsilon$ (recall that $d=3-\epsilon$). However, the integral is actually finite as $\epsilon \to 0$; the reason is that for large enough $\vec k$ the two poles for $\omega$ are on the same side of the complex plane and the integral over $\omega$ vanishes. The final result is (see the Appendix for more details)
\be\label{eq:Pifinite}
\Pi(p_0, \vec p)= \frac{g^2 k_F^2}{2\pi^2 vN} \left(1- \frac{p_0}{v |\vec p|}\,{\rm \tan}^{-1}\left(\frac{v | \vec p|}{p_0}\right) \right)\,.
\ee

As $\epsilon \to 0$ we find a finite answer and no pole in $\epsilon$ --the fermion loop does not contribute to the anomalous dimension of the boson. Nevertheless, the finite contribution is nontrivial: it is the familiar Landau damping term (usually calculated in the patch description), now evaluated in the `spherical' RG towards the Fermi surface. At small frequency and momenta and fixed $v$, (\ref{eq:Pifinite}) is more relevant than the tree level kinetic term. This means that the Wilson-Fisher type fixed point is a good approximation over a finite range of energy and momenta, outside of which quantum corrections to the kinetic term cannot be neglected.
However, as first noted in~\cite{shamitfuture}, the running of the velocity $v \to 0$ that we found before becomes important here. Indeed, at fixed frequency and momenta but small $v$, the Landau damping contribution behaves as
\be
\Pi(p_0, \vec p) \approx \frac{g^2 k_F^2}{6\pi^2 N}\,\frac{|\vec p|^2}{p_0^2}\,v\,.
\ee
Therefore, if the scale at which (\ref{eq:Pifinite}) becomes important is smaller than the RG scale at which the velocity reaches its fixed point $v=0$, the corrections from Landau damping to the critical scalar will be negligible. This can be accomplished by a suitable choice of boundary conditions for the couplings $\lambda_0$ and $g_0$.

The remaining effect of fermions is the box diagram that renormalizes the $\phi^4$ coupling. For vanishing external momenta, this contribution is given by
\be
\Gamma^{(4)}(p_i)= - 144 \frac{g^4}{N^2}\,\delta^4(\sum p_i)\,\int \frac{d\omega d^d k}{(2\pi)^{d+1}}\,\prod_{i=1}^4\frac{1}{i (\omega+\omega_i) - \frac{(\vec k+\vec p_i)^2}{2m}+\mu_F}\,.
\ee
The integral is convergent in dimensional regularization and is calculated explicitly in the Appendix, with the result that $\Gamma^{(4)} \to 0$ as $p_i \to 0$. This correction corresponds to an irrelevant operator and can be safely neglected at low energies. 

We conclude that the approximate critical theory for the scalar field survives at finite $N$. The fixed point is valid up to the scale where Landau damping becomes relevant. Furthermore, for appropriate choices of the UV couplings the fermion speed reaches $v=0$ before the Landau damping scale, and then the fixed point is valid over the original exponential window (\ref{eq:windowkappa1}).

\subsection{Effects on the non Fermi liquid}\label{subsec:NFL-finiteN}

At finite $N$ there are two types of quantum corrections that could affect the non Fermi liquid.

First, the bosons mediate a $\psi^4$ interaction which is attractive and induces a Cooper pairing instability. Since the theory is perturbative, the superconducting gap $\Delta$ is exponentially small in the $\psi^4$ coupling, as in Fermi liquid theory. This destroys the non Fermi liquid phase at a scale that is exponentially small in the Yukawa coupling.\footnote{In fact, as we discussed before, at exponentially small scales the non-Fermi liquid is also affected by the gradients (\ref{eq:gradients1}).} A superconducting instability may also be a feature instead of a problem, since many materials that display non Fermi liquid behavior also become superconducting at small temperatures.

Secondly, at finite $N$ we need to take into account corrections to the Yukawa vertex $\phi \bar \psi \psi$, which we will denote by $\Gamma^{(3)}$. Let us set $N=1$ for now, and we will determine the appropriate group theory factors below. Expanding in $1/k_F$, the three-point function with incoming fermion
momentum $k$ and outgoing boson momentum $q$ is
\bea
\Gamma^{(3)}\left(k_{0},k_{\perp},q_{0},q_{\perp}\right)  &=&  \frac{g^{3}}{\left(2\pi\right)^{d+1}}\int d\omega d\ell d^{d-1}p \frac{1}{\omega^{2}+c^{2}\left(\ell^{2}+p^{2}\right)}
\frac{1}{i\left(\omega+k_{0}\right)-v\left(\ell+k_{\perp}\right)}\times \nonumber\\
&& \qquad \qquad \times\frac{1}{i\left(\omega+k_{0}-q_{0}\right)-v\left(\ell+k_{\perp}-q_{\perp}\right)}\,.
\eea
Fortunately we will not need to evaluate this integral explicitly, because for $q_{0}=q_{\perp}=0$ it can be related to the integral
for the fermion self-energy:
\be\label{eq:Gamma3}
\Gamma^{(3)}\left(\omega,\ell,0,0\right) =  -ig\frac{d}{d\omega}\Sigma\left(\omega,\ell\right)= \frac{g^3}{4\pi^2 c^2(c+\left|v\right|)} \frac{1}{\epsilon} + \ldots
\ee
We should note that, similarly to what was observed for $\Gamma^{(4)}$ in the Appendix, the limits $q_0 \to 0$ and $q_\perp \to 0$ do not commute. Here by (\ref{eq:Gamma3}) we mean $q_\perp \to 0$ first and $q_0$ is kept finite for the evaluation of the loop integral (\ref{eq:Gamma3}). In a theory with a gauge field instead of a scalar boson, this order of limits turns out to be required by the gauge theory Ward identity~\cite{future}.

Going back to the renormalized action (\ref{eq:Lrenorm}), the divergent vertex (\ref{eq:Gamma3}) means that at finite $N$ we also need a nonzero counterterm $\delta_g= \delta_Z$. Note that for the simple example of a single scalar interacting with a fermion ($N=1$), the one loop contributions of $\delta_g$ and $\delta_Z$ to the beta function of $g$ [Eq.~(\ref{eq:RG1})] precisely cancel,\footnote{Similar (but exact to all orders) cancellations occur in QED, where the Ward identity relates the vertex and wavefunction renormalizations, and the gauge coupling running depends only on the photon vacuum polarization.} obtaining
\be
\beta_g =g( - \kappa_2  +  \gamma_\phi) \,,
\ee
and $\gamma_\phi=0$ at one loop.
In this case we have a non Fermi liquid which is no longer critical, because the Yukawa coupling is classically relevant and its beta function is nonzero along the flow.

Next, let us consider a model with an $SU(N)$ global symmetry, with $\psi$ in the fundamental and $\phi$ in the adjoint representations. Denote the generators of $SU(N)$ in the fundamental representation $\Box$ by $T^a$, normalized to ${\rm tr}(T^a T^b)=\delta^{ab}$.\footnote{We choose this normalization to agree with our previous large $N$ conventions.} The fermion self-energy is now proportional to $T^a T^a = C_2({\rm \Box})=\frac{N^2-1}{N}$, while the group-theoretic factor for the vertex is $T^b T^a T^b=(C_2({\rm \Box})-\frac{1}{2}C_2({\rm adj})) T^a$, with $C_2({\rm adj})=2N$. The counterterms then read
\be
\delta_Z = - \frac{g^2}{4\pi^2 c^2 (c+\left|v\right|)}\,\frac{C_2({\rm \Box})}{N}\, \frac{1}{\epsilon}\;,\;\delta_g=-\frac{g^2}{4\pi^2 c^2 (c+\left|v\right|)}\,\frac{C_2({\rm \Box})-\frac{1}{2}C_2({\rm adj})}{N}\,\frac{1}{\epsilon}\,.
\ee
Plugging these results into the expressions for the anomalous dimension and beta functions gives
\bea
\gamma&=& \frac{g^2}{8 \pi^2 c^2(c+\left|v\right|)}\,\frac{C_2({\rm \Box})}{N} \nonumber\\
\beta_v&=&\frac{g^2}{4 \pi^2 c^2}\,\frac{C_2({\rm \Box})}{N}\,{\rm sign}(v)\\
\beta_g&=& g \left(- \frac{\epsilon}{2}- \kappa_2+ \frac{g^2}{4\pi^2c^2(c+|v|)}\,\frac{C_2({\rm adj})}{2N} \right)\,, \nonumber
\eea
which extend (\ref{eq:RG1}) to finite $N$. In particular, in the abelian case $C_2({\rm adj})=0$, and we recover the cancellation between $\delta_Z$ and $\delta_g$ in the beta function.

At finite $N$ we then find a critical non Fermi liquid described by $v \to 0$, a perturbative fixed point for the Yukawa interaction
\be
g^2 \approx 4\pi^2 c^3 \kappa_2
\ee
(which is the same as in the large $N$ limit), and an anomalous dimension for the fermion
\be
\gamma \approx \left(1-\frac{1}{N^2} \right)\frac{\kappa_2}{2}\,.
\ee

\section{Conclusions}\label{sec:concl}

In this work we have shown that mild spacetime dependent gradients can lead to perturbatively controlled non-Fermi liquids in $3+1$ spacetime dimensions. The theory contains nonrelativistic fermions at finite density interacting with a critical boson via a Yukawa coupling, $g(x) \phi \psi^\dag \psi$. The non-Fermi liquid behavior is obtained by balancing the classical spacetime dependence of $g(x)=g_0|x|^\kappa$ (with $0<\kappa \ll 1$) against one loop effects, and it is valid over an exponentially large but finite range of scales. We also exhibited an approximate fixed point for the boson by adding a $\lambda(x)\phi^4$ self-interaction with a similar type of gradient, as in~\cite{Dong:2012ua}. At one loop the running of the Yukawa and $\phi^4$ couplings are independent --in particular, the finite density fermions do not contribute to $\beta_\lambda$. A special and simpler case of our mechanism is then $\lambda(x)=0$, for which there is still a non-Fermi liquid behavior induced by the Yukawa interaction alone.

An important direction of research would be to understand whether gradients can provide control in $2+1$ dimensions, where the theory of a critical boson interacting with a Fermi surface is believed to be responsible for the non-Fermi liquid behavior observed in strongly correlated materials. It is still possible to introduce translation breaking so as to render the Yukawa coupling nearly marginal classically, but in 2+1 dimensions this would require order one values of $\kappa$, going beyond the perturbative treatment of the translation breaking that we have used in this paper.  Such a regime of strong translation breaking may still be consistent with weak Yukawa coupling, as some preliminary analysis suggests \cite{unpublished}.  It would be interesting to relate this to other work such as \cite{Seantransl}\ classifying the low energy behavior of translation breaking effects. 

 Another possibility would be to try to combine the effects of gradients and the large $N$ expansion in order to obtain a controlled non-Fermi liquid in 2+1 dimensions. In particular, one could try to use spacetime dependent couplings to suppress the effects found in~\cite{lee1} that invalidate the large $N$ limit of~\cite{Polchinski:1993ii} . Finally, given the very minimal matter content of the theory, it would be interesting to think about possible experimental realizations of our mechanism.

\section*{Acknowledgements}

We thank M. Dodelson, L. Fitzpatrick, S. Hartnoll, S. Kachru, S. Kivelson, S.-S. Lee, H. Liu, J. McGreevy, S. Raghu and H. Wang for useful discussions.  
E.S. is grateful to the organizers and participants of the 2013 Simons Symposium on ``Quantum Entanglement'' for useful feedback. S.M. was supported in part by an award from the Department of Energy (DOE) Office of Science Graduate Fellowship Program.  This research program was made possible in part by the National Science Foundation under grant PHY-0756174, by the Department of Energy under contract DE-AC03-76SF00515, and by readers like you.

\appendix

\section{One loop calculations in dimensional regularization}

\subsection{Fermion self-energy}

The calculation of the fermion self-energy at finite density is somewhat involved, and the form of the finite terms is important for the RG of the theory. So in this Appendix we perform explicitly this calculation (using dimensional regularization) and reproduce the results in~\cite{NFLepsilon}, up to a typo in a relative sign.  We give the result to leading order in $\ell/k_F$.

The integral is:
\be
\Sigma  =  \frac{-g^{2} \mu^\epsilon}{\left(2\pi\right)^{d+1}}\times\int d\omega d\ell d^{d-1}k\frac{1}{\omega^{2}+c^{2}\left(\ell^{2}+k^{2}\right)}\frac{1}{i\left(\omega+\omega_{e}\right)-v\left(\ell+\ell_{e}\right)}\,.
\ee
We first do the $k$ integral, whose general form is:
\be
\int d^{d-1}k\frac{1}{x+k^{2}}  =  S_{d-2}\int dk\frac{k^{d-2}}{x+k^{2}}= \frac{\pi S_{d-2}}{2\sin\left(\frac{1}{2}\pi\left(d-1\right)\right)}x^{\left(d-3\right)/2}\,,
\ee
where $S_{d-1} = 2 \pi^{d/2}/\Gamma\left(d/2\right)$ is the area of a $\left(d-1\right)$-sphere. This gives us, in terms of $d=3-\epsilon$,
\be
\Sigma=\frac{g^{2}\mu^\epsilon S_{d-2}}{\left(2\pi\right)^{d+1}}\frac{\pi}{2c^{2-\epsilon}\sin\left(\frac{\pi\epsilon}{2}\right)}\times\int d\omega d\ell\frac{1}{\left(\omega^{2}+c^{2}\ell^{2}\right)^{\epsilon/2}}\frac{i\left(\omega+\omega_{e}\right)+v\left(\ell+\ell_{e}\right)}{\left(\omega+\omega_{e}\right)^{2}+v^{2}\left(\ell+\ell_{e}\right)^{2}}\,.
\ee
Using Feynman parameters
\be
\frac{1}{A_{1}^{m_{1}}A_{2}^{m_{2}}\ldots A_{n}^{m_{n}}}=\int_{0}^{1}dx_{1}\ldots dx_{n}\delta\left(\sum x_{i}-1\right)\frac{\Pi x_{i}^{m_{i}-1}}{\left(\sum x_{i}A_{i}\right)^{\sum m_{i}}}\frac{\Gamma\left(\sum m_{i}\right)}{\Pi\Gamma\left(m_{i}\right)}
\ee
then gives
\begin{eqnarray*}
\Sigma & = & \frac{g^{2} \mu^\epsilon S_{d-2}}{\left(2\pi\right)^{d+1}}\frac{\pi}{2c^{2-\epsilon}\sin\left(\frac{\pi\epsilon}{2}\right)}\times\frac{\Gamma\left(1+\frac{\epsilon}{2}\right)}{\Gamma\left(\frac{\epsilon}{2}\right)}\int d\omega d\ell\,\times\\
&& \qquad \times \int dxdy\delta\left(x+y-1\right)y^{\epsilon/2-1}\frac{i\left(\omega+\omega_{e}\right)+v\left(\ell+\ell_{e}\right)}{\left(y\left(\omega^{2}+c^{2}\ell^{2}\right)+x\left(\left(\omega+\omega_{e}\right)^{2}+v^{2}\left(\ell+\ell_{e}\right)^{2}\right)\right)^{1+\epsilon/2}}\\
 & = & \frac{g^{2} \mu^\epsilon S_{d-2}}{\left(2\pi\right)^{d+1}}\frac{\pi}{2c^{2-\epsilon}\sin\left(\frac{\pi\epsilon}{2}\right)}\times\frac{\epsilon}{2}\times\int dx\left(1-x\right)^{\epsilon/2-1}\times I_{1}
\end{eqnarray*}

The required integral now is:
\be
I_{1}=\int d\omega d\ell\frac{i\left(\omega+\omega_{e}\right)+v\left(\ell+\ell_{e}\right)}{\left(\omega^{2}+\left(xv^{2}+c^{2}\left(1-x\right)\right)\ell^{2}+x\left(2\omega\omega_{e}+2v^{2}\ell\ell_{e}+\omega_{e}^{2}+v^{2}\ell_{e}^{2}\right)\right)^{1+\epsilon/2}}
\ee
We can shift $\omega\rightarrow\omega-\omega_{e}x$ and $\ell\rightarrow\ell-\frac{v^{2}\ell_{e}}{\left(xv^{2}+c^{2}\left(1-x\right)\right)}x$
to complete the square in the denominator:
\be
I_{1}=\int d\omega d\ell\frac{i\left(\omega+\omega_{e}\left(1-x\right)\right)+v\left(\ell+\ell_{e}\left(1-\frac{v^{2}x}{\left(xv^{2}+c^{2}\left(1-x\right)\right)}\right)\right)}{\left(\omega^{2}+\left(xv^{2}+c^{2}\left(1-x\right)\right)\ell^{2}-\frac{\left(v^{2}\ell_{e}x\right)^{2}}{\left(xv^{2}+c^{2}\left(1-x\right)\right)}+xv^{2}\ell_{e}^{2}+x\left(1-x\right)\omega_{e}^{2}\right)^{1+\epsilon/2}}
\ee
We can discard the $\omega$ and $\ell$ terms in the numerator, because
the denominator is even in both:
\be
I_{1}=\int d\omega d\ell\frac{i\omega_{e}\left(1-x\right)+v\ell_{e}\left(1-\frac{v^{2}x}{\left(xv^{2}+c^{2}\left(1-x\right)\right)}\right)}{\left(\omega^{2}+\left(xv^{2}+c^{2}\left(1-x\right)\right)\ell^{2}-\frac{\left(v^{2}\ell_{e}x\right)^{2}}{\left(xv^{2}+c^{2}\left(1-x\right)\right)}+xv^{2}\ell_{e}^{2}+x\left(1-x\right)\omega_{e}^{2}\right)^{1+\epsilon/2}}
\ee
 The general integral we need is:
\be
\int\frac{dadb}{\left(a^{2}+cb^{2}+y\right)^{z}}=\frac{\pi y^{1-z}}{c^{1/2}\left(z-1\right)}
\ee
Then, we have:
\begin{eqnarray*}
I_{1} & = & \left(i\omega_{e}\left(1-x\right)+v\ell_{e}\left(1-\frac{v^{2}x}{\left(xv^{2}+c^{2}\left(1-x\right)\right)}\right)\right)\times\\
&& \qquad \times \frac{\pi}{\left(xv^{2}+c^{2}\left(1-x\right)\right)^{1/2}\left(\frac{\epsilon}{2}\right)}\left(xv^{2}\ell_{e}^{2}-\frac{\left(v^{2}\ell_{e}x\right)^{2}}{\left(xv^{2}+c^{2}\left(1-x\right)\right)}+x\left(1-x\right)\omega_{e}^{2}\right)^{-\epsilon/2}\\
 & = & \frac{2\pi}{\epsilon}\frac{\left(1-x\right)^{1-\epsilon/2}\left(i\omega_{e}\left(xv^{2}+c^{2}\left(1-x\right)\right)+v\ell_{e}c^{2}\right)}{\left(xv^{2}+c^{2}\left(1-x\right)\right)^{\left(3-\epsilon\right)/2}\left(v^{2}\ell_{e}^{2}c^{2}+x\omega_{e}^{2}v^{2}+c^{2}\omega_{e}^{2}\left(1-x\right)\right)^{\epsilon/2}x^{\epsilon/2}}
\end{eqnarray*}

Inserting this result into $\Sigma$, we arrive to
\be
\Sigma=\frac{g^{2}S_{d-2}}{\left(2\pi\right)^{d+1}}\frac{\pi^{2}}{2c^{2-\epsilon}\sin\left(\frac{\pi\epsilon}{2}\right)}\int dx\frac{i\omega_{e}\left(xv^{2}+c^{2}\left(1-x\right)\right)+v\ell_{e}c^{2}}{\left(xv^{2}+c^{2}\left(1-x\right)\right)^{\left(3-\epsilon\right)/2}\left(v^{2}c^2\frac{\ell_{e}^{2}}{\mu^2}+xv^2\frac{\omega_{e}^{2}}{\mu^2}+c^{2}\left(1-x\right)\frac{\omega_{e}^{2}}{\mu^2}\right)^{\epsilon/2}x^{\epsilon/2}}
\ee
Expanding in $\epsilon$ gives 
\be
\Sigma=\frac{\Sigma_{-1}}{\epsilon}+\Sigma_{0}+\mathcal O\left(\epsilon\right).
\ee
with
\begin{eqnarray*}
\Sigma_{-1} & = & \left(\frac{g^{2}2\pi}{\left(2\pi\right)^{4}}\right)\frac{\pi}{c^{2}}\int dx\left(\frac{i\omega_{e}}{\left(xv^{2}+c^{2}\left(1-x\right)\right)^{1/2}}+\frac{v\ell_{e}c^{2}}{\left(xv^{2}+c^{2}\left(1-x\right)\right)^{3/2}}\right)\\
 & = & \frac{g^{2}}{4\pi^{2}c^{2}\left(c+\left|v\right|\right)}\left(i\omega_{e}+\text{sgn}\left(v\right)c\ell_{e}\right)
\end{eqnarray*}
and
\begin{eqnarray*}
\Sigma_{0} & = & \left(\frac{g^{2}}{16\pi^{2}c^{2}}\right)\int dx\left(\left\{ 2\log\left(2\pi c\right)-\gamma-\log\pi\right\} +\log\left(\frac{xv^{2}\mu^2+c^{2}\left(1-x\right)\mu^2}{x\left(\omega_{e}^{2}\left(xv^{2}+\left(1-x\right)c^{2}\right)+v^{2}\ell_{e}^{2}c^{2}\right)}\right)\right)\times \\
&& \qquad \qquad \times\,\frac{i\omega_{e}\left(xv^{2}+c^{2}\left(1-x\right)\right)+v\ell_{e}c^{2}}{\left(xv^{2}+c^{2}\left(1-x\right)\right)^{3/2}}\\
 & = & \left\{ 2\log\left(2\pi c\right)-\gamma-\log\pi\right\} \times\Sigma_{-1}-\frac{g^{2}}{16\pi^{2}c^{2}}\int dx\,\left\{ \log x+\log\left(\frac{\omega_{e}^{2}}{\mu^2}+\frac{\ell_{e}^{2}v^{2}c^{2}}{xv^{2}\mu^2+\left(1-x\right)c^{2}\mu^2}\right)\right\}\times \\
 && \qquad \qquad \times\, \left\{ \frac{i\omega_{e}}{\left(xv^{2}+c^{2}\left(1-x\right)\right)^{1/2}}+\frac{v\ell_{e}c^{2}}{\left(xv^{2}+c^{2}\left(1-x\right)\right)^{3/2}}\right\} \,.
\end{eqnarray*}
Evaluating the integral in the last line gives
\begin{eqnarray*}
\Sigma_{0}
 & = & \frac{g^{2}}{8\pi^{2}c^{2}\left(c^{2}-v^{2}\right)}\bigg(2c\left(v\ell_{e}-i\omega_{e}\right)\log\left(\frac{2\left(v\ell_{e}-i\omega_{e}\right)}{(c+v)\mu}\right)-(c+v)\left(c\ell_{e}-i\omega_{e}\right)\log\left(\frac{c\ell_{e}-i\omega_{e}}{c \mu}\right)\nonumber\\
 &&-(c-v)\left(c\ell_{e}+i\omega_{e}\right)\log\left(\frac{c\ell_{e}+i\omega_{e}}{c\mu}\right)+\left\{ 2c\left(v\ell_{e}-i\omega_{e}\right)-(c+v)\left(c\ell_{e}-i\omega_{e}\right)\right\} \log\left(c\right)\bigg)\\
 &  & +\left(\log\left(4\pi\right)-\gamma+4\right)\Sigma_{-1}
\end{eqnarray*}

\subsection{Boson self-energy}

The one-loop boson self-energy with finite density fermions running in the loop is given by
\be
\Pi(p_0, \vec p)= g^2 \int \frac{d\omega\,d^d k}{(2\pi)^{d+1}} \,\frac{1}{i \omega - \frac{\vec k^2}{2m}+\mu_F}\, \frac{1}{i (\omega+p_0) - \frac{(\vec k+\vec p)^2}{2m}+\mu_F}\,.
\ee
Integrating over $\omega$ using residues gives
\be
\Pi(p_0, \vec p)= i g^2 \int \frac{d^d k}{(2\pi)^{d}} \,\frac{{\rm sign}(\mu_F - \frac{\vec k^2}{2m})\Theta\left({\rm sign}(\mu_F - \frac{\vec k^2}{2m}) \left[\frac{(\vec k+ \vec p)^2}{2m}-\mu_F\right]\right)}{p_0+\frac{i}{2m} \left(2 \vec k \cdot \vec p + \vec p^2 \right)}\,.
\ee
The Heaviside function comes from requiring that the two poles be on opposite sides of the real axis.

Note that the step function makes the integral convergent, so we can set $d=3$, and it is enough to expand around the Fermi surface,
\be
\vec k = (k_F + l) \hat n\;,\; \frac{\vec k^2}{2m}-\mu_F \approx v l\;,\;\frac{(\vec k+\vec p)^2}{2m}-\mu_F \approx v( l + \hat n \cdot \vec p)\,,
\ee
with $\hat n$ a unit vector.
The integral then simplifies to
\be
\Pi(p)= \frac{i g^2 k_F^2}{4\pi^2}\,\int_{-1}^1 d(\cos \theta)\,\int dl\,{\rm sign}l\,\frac{\Theta\left(-{\rm sign}l (l+  |\vec p| \cos \theta) \right)}{p_0+ i v | \vec p| \cos \theta}\,.
\ee
Performing the remaining integrals gives
\be
\Pi(p_0, \vec p)= \frac{g^2 k_F^2}{2\pi^2 v} \left(1- \frac{p_0}{v |\vec p|}\,{\rm \tan}^{-1}\left(\frac{v | \vec p|}{p_0}\right) \right)\,.
\ee

\subsection{Scalar $\phi^4$ coupling}

As with the one-loop correction to the Yukawa coupling (\ref{eq:Gamma3}), the one-loop correction to the $\phi^4$ interaction due to the fermions can be related to the boson self-energy:
\begin{eqnarray}
\Gamma^{\left(4\right)}\left(0,0,0\right)&=&-\frac{g^{2}}{2}\frac{\partial^{2}}{\partial p_{0}^{2}}\Pi\left(p_{0},\vec{p}\right)\big|_{p_{0},\left|p\right|=0}\nonumber\\
&=&-\frac{g^{4}k_{F}^{2}}{2\pi^{2}v}\frac{v_{F}^{2}\left|p\right|^{2}}{\left(v_{F}^{2}\left|p\right|^{2}+p_{0}^{2}\right)^{2}}\big|_{p_{0},\left|p\right|=0}
\end{eqnarray}
This indeed vanishes when we hold $p_0$ finite as we take $|\vec{p}| \rightarrow 0$.  However, we cannot check the other limit, because the $p_0$ derivative demands that we hold $p_0$ finite.  In fact, when one naively takes $p_0 \rightarrow 0$ before $\vec{p} \rightarrow 0$, this diverges.  Hence, we will need to compute this correction directly and resolve the ambiguity of limits.

The diagram at arbitrary external momentum is given by:
\begin{eqnarray}
\Gamma^{\left(4\right)}\left(p,q,r\right) & = & g^{4}\int\frac{d\omega d^{d}k}{\left(2\pi\right)^{d+1}}~G_{F}\left(\omega,k\right)G_{F}\left(\omega+p_{0},k+p\right)\times\nonumber\\
 &  & \times G_{F}\left(\omega+q_{0},k+q\right)G_{F}\left(\omega+r_{0},k+r\right)
\end{eqnarray}
where the external momenta $p,q-p,r-q,-r$ are all taken much less
than $k_{F}$. Noting that the integral will be dominated by $\vec{k}$
close to the Fermi surface, set $\vec{k}=\hat{n}\left(k_{F}+\ell\right)$
and expand $\frac{\left(k+q\right)^{2}}{2m}-\mu_{F} \approx v\left(\ell+\hat{n}\cdot q\right)$:
\begin{eqnarray}
\Gamma^{\left(4\right)}\left(p,q,r\right) & = & \frac{g^{4}k_{F}^{2}}{\left(2\pi\right)^{4}}\int_{-k_{F}}^{\infty}d\ell\int d\omega d^{2}\hat{n}~\frac{1}{i\omega-v\ell}\frac{1}{i\left(\omega+p_{0}\right)-v_{F}\left(\ell+\hat{n}\cdot p\right)}\times\nonumber\\
 &  & \times\frac{1}{i\left(\omega+q_{0}\right)-v_{F}\left(\ell+\hat{n}\cdot q\right)}\frac{1}{i\left(\omega+r_{0}\right)-v_{F}\left(\ell+\hat{n}\cdot r\right)}
\end{eqnarray}
After evaluating the $\omega$ contour integral, the $\ell$ integral
is finite, and we obtain:
\begin{eqnarray}
\Gamma^{\left(4\right)}\left(p,q,r\right) & = & \frac{g^{4}k_{F}^{2}}{\left(2\pi\right)^{4}}\int d^{2}\hat{n}\times\\
 &  & \times\frac{\hat{n}\cdot p}{\left(p_{0}+iv_{F}\hat{n}\cdot p\right)\left(\left(q_{0}-p_{0}\right)+iv_{F}\hat{n}\cdot\left(q-p\right)\right)\left(\left(r_{0}-p_{0}\right)+iv_{F}\hat{n}\cdot\left(r-p\right)\right)}+\text{cyc.}\nonumber
\end{eqnarray}
where ``cyc.'' denotes cyclic permutations of $p,q,r$.

Now, setting $\vec{p}=\vec{q}=\vec{r}=0$ while keeping $p_{0},q_{0},r_{0}$
constant, the integrand is trivially zero, confirming our result above.
In addition, setting $p_{0}=q_{0}=r_{0}=0$ while keeping $\vec{p},\vec{q},\vec{r}$
constant, the cyclic sum is zero: 
\begin{eqnarray}
\Gamma^{\left(4\right)}\left(p,q,r;p_{0}=q_{0}=r_{0}=0\right) & = & \frac{ik_{F}^{2}}{v_{F}^{3}\left(2\pi\right)^{4}}\int d^{2}\hat{n}\left\{ \frac{1}{\hat{n}\cdot\left(q-p\right)}\frac{1}{\hat{n}\cdot\left(r-p\right)}+\text{cyc.}\right\} \nonumber\\
 & = & 0
\end{eqnarray}
Hence, we have $\Gamma^{\left(4\right)}\left(0,0,0\right)=0$, and
there is no unexpected effect of the fermions on the bosons at small
momentum.

\bibliographystyle{JHEP}
\renewcommand{\refname}{Bibliography}
\addcontentsline{toc}{section}{Bibliography}
\providecommand{\href}[2]{#2}\begingroup\raggedright

\end{document}